\documentclass[11pt]{llncs}

\usepackage{graphicx}
\usepackage{subfig}
\usepackage{amssymb}
\usepackage{booktabs}
\begin{document}

\title{Anonymity in the wild: Mixes on unstructured networks}
\author{Shishir Nagaraja}
\institute{Computer Laboratory\\
JJ Thomson Avenue, Cambridge CB3 0FD, UK\\
{\tt shishir.nagaraja@cl.cam.ac.uk}}
\maketitle

\begin{abstract}
As decentralized computing scenarios get ever more popular,
unstructured topologies are natural candidates to consider running mix
networks upon. We consider mix network topologies where mixes are
placed on the nodes of an unstructured network, such as social
networks and scale-free random networks. We explore the
efficiency and traffic analysis resistance properties of mix networks
based on unstructured topologies as opposed to theoretically optimal
structured topologies, under high latency conditions. We consider a
mix of directed and undirected network models, as well as one real
world case study -- the LiveJournal friendship network topology. Our
analysis indicates that mix-networks based on scale-free and
small-world topologies have, firstly, mix-route lengths that are
roughly comparable to those in expander graphs; second, that
compromise of the most central nodes has little effect on
anonymization properties, and third, batch sizes required for warding
off intersection attacks need to be an order of magnitude higher in
unstructured networks in comparison with expander graph topologies.
\end{abstract}

\section{Introduction}
As governments pursue large scale surveillance and censorship
programs, anonymity in online communication mechanisms is an
increasingly important requirement. Anonymous communications are also
useful in building resistance against a global passive adversary who
can subject the targets to traffic analysis. Often, an attacker will
try to destabilize a network by building a dossier of the most central
nodes, and attacking ones on the top of the list. Traffic analysis of
inter-node communication offers basic tools to collect necessary
intelligence in order to plan an attack.

Seminal work by Chaum~\cite{C81} introduced mix networks as a
technique to provide anonymous communications where messages are
relayed through a sequence of intermediate nodes called mixes, to make
the task of tracing them through the network as difficult as
possible. The essential idea is to make the inputs of each mix bit-wise
unlinkable to its outputs.

Anonymity research conducted since, can be classified into low-latency
or real time systems primarily for Internet browsing such as
onion routing~\cite{STRL00} and high-latency or non-real time systems
such as mixminion~\cite{DDM03}.

The topology of a mix network plays an important role in its
efficiency and traffic analysis resistance properties. The mainstream
design paradigm that has emerged so far is that of structured network
topologies based on regular graphs. The theory is that such topologies
are amenable to theoretical analysis that proves they have optimal
expansion properties. This leads to a mix network design that is
highly efficient and resistant to traffic analysis. Examples are
onion-routing systems such as TOR~\cite{DMS04}that use a complete
graph topology, where a mix can contact every other mix in the
network. While such models are theoretically elegant, the assumption
that every node in the network is equally resourced (as regular graphs
necessitate) to handle network traffic loads is their main drawback.

An alternate paradigm is topology based on unstructured networks, such
as those inspired from social networks. The argument in their favor
being that the incentive to carry traffic is clear and simple -
friends carry each-others traffic. Moreover, no additional resources
go into constructing an overlay network since the pre-existing
topology is used by the mix network as well, which works well for
power constrained environments such as adhoc networks and sensor
networks. Legal considerations play an important role too. It is not
enough to merely have a large number of mixes. When hassled by legal
requests (such as a subpoena to hand-over mix server logs to the
police), a mix-network where friends route each others traffic, is
likely to have a higher proportion of servers in operation, as opposed
to a synthetic network.

%Individuals and groups have varying requirements of anonymity. Some
%people merely want to surf the Web, purchase online, and send email
%without exposing to marketing agencies what their interests and
%identities are; Political dissidents and opposition groups need to
%protect themselves against domestic surveillance, protest groups need
%to make sure their leaders cannot be identified and taken away in a
%police van; corporations must communicate with various agencies
%without exposing the existence of communications to competitors; and
%military units must communicate with each-other without revealing field
%intelligence. Anonymity on the Internet, is thus a requirement for
%many users with varying levels of sensitivity and awareness to
%disclosure. See \cite{DM06} for an extended discussion.

%Unlike other communication security primitives such as confidentiality
%(encryption functions) and message integrity (cryptographic hash
%functions), anonymity cannot be created by the user. Mixes work by
%'losing' messages into the network, for which you need heavy
%traffic. Anonymity systems use messages to hide messages by mixing
%them together, senders of messages use the anonymity the mix network
%infrastructure provides, while providing cover traffic that in turn
%drives the system. In order to make a phone call a drug dealer might
%choose a road clogged with rush hour traffic rather over a quiet
%neighbourhood where her location might be given away by background
%sounds.

A comparison between the two paradigms needs to address mix-network
efficiency, resilience to corrupt nodes and the loss of anonymity from
statistical disclosure attacks.

In this paper we analyze various types of unstructured networks,
especially social networks and evaluate their suitability as mix
topologies. We discuss the reasons behind using social networks to
route mix traffic and we analyze the suitability of various types of
model networks to routing mix traffic and offer a comparison between
them. We also analyze the theoretical bounds on anonymity such
networks can provide in terms of mixing speed and resistance to
traffic analysis. We apply concepts from spectral graph theory to
derive the route length necessary to provide maximal anonymity.

This paper is organized as follows: Section 3 discusses the various
topologies used in our analysis. Section 4, lays out the evaluation
framework to measure the traffic analysis resistance of various
topologies. Section 5 discusses the application of the framework to
various topologies and the results obtained. Finally, we offer our
conclusions in section 6.

\section{Related work}
Danezis~\cite{D03} explored the anonymity provided by expander graph
topologies, this is one of the main sources of inspiration for our work. He
established the thoretical bounds of anonymity for expander graphs,
and also showed that they were optimal.

Borisov~\cite{B05} analyzes anonymous communications over a De Bruijn
graph topology overlay network. He analyzes the deBruijn graph
topology and comments on their successful mixing capabilities.

\section{Network models}
\label{sec-netmod}
In this section we give a brief introduction to the network models we
wish to analyze as candidates for mix network topologies.

\subsection{Erd\"{o}s-R\'{e}nyi model of random networks}
On the earliest models for heterogeneous networks is the
Erd\"{o}s-R\'{e}nyi (ER )model~\cite{ER59}. Although seldom found in
real world networks, their use has been popularised by the work of
Eschenauer and Gligor~\cite{EG02} is designing a key management scheme
for sensor networks.

Here, we start from $N$ vertices without any
edges. Subsequently, edges connecting two randomly chosen vertices are
added as the result of a Bernoulli trial, with a parameter $p$. It
generates random networks with no particular structural bias. The
average degree $\langle k \rangle = 2L/N$ where $L$ is the total number of edges, can
also be used as a control parameter. ER model networks have a
logarithmically increasing $l$, a normal degree distribution, and a
clustering coefficient close to zero.

%We used $N$=5000 nodes, $<k>$=14, and $p$=0.0014. The analysis of
%mixing rates is at the end of the next subsection~\ref{sec-sfba}.

\subsection{Scale-free networks with linear preferential attachment}
\label{sec-sfba}
A number of popular peer-to-peer systems are found to have
heterogeneous topologies with heavy tailed degree distributions. The
work of Ripeanu~\cite{RFI02} shows that two popular systems,
Gnutella~\cite{KM02} and Freenet~\cite{CSWH00}, have power-law degree
distributions.

A variable X is said to follow a heavy tail distribution if \(Pr[X>x]
\sim x^{-k}\ L(x)\) where \(k \in \Re^{+}\) and $L(x)$ is a slowly
varying function so that
\(\lim_{x\to\infty}{\frac{L(tx)}{L(x)}}\to1\). A power-law
distribution is simply a variation of the above where one studies
\(Pr[X=x] \sim x^{-(k+1)}=x^{-\alpha}\). The degree of a node is the
number of links it has to other nodes in the network. If the degree
distribution of a network follows a power-law distribution it is known
as a scale-free network. The power-law in the degree or link
distribution reflects the presence of central individuals who interact
with many others on a continual basis and play a key role in relaying
information.

We denote a scale-free network generated by preferential attachment,
by $G_{m, N}(V, E)$ where $m$ is the number of initial nodes created
at time=$t_{0}$ and $N$ is the total number of nodes in the
network. At every time step $t_{i}, i \ge 0$, one node is added to the
network. For every node $v$ added, we create $m$ edges from the $v$ to
existing nodes in the network according to the following linear
preferential attachment function due to Barabasi and Albert~\cite{AB02}:
\[Pr[(v,i)] = k_{i}/\sum_{j}k_{j}\] where $k_{i}$ is the degree of node $i$. 
We continue until $|V| = N$.

\subsection{Scale-free random graph topology}
\label{sec-sfrgt}
An alternate way of constructing a large scale-free network is to
create a network with a given power-law degree sequence that is random
in all other aspects. Aiello et.al.~\cite{ACL00} propose such a random
graph model inspired by massive AT\&T call graphs, with two parameters
$\alpha$ and $\beta$. Where, $\alpha$ gives the fraction of
nodes with degree 1 and $\beta$ defines the exponent of the power-law
function. Then, if $y$ be the number of vertices of degree $x>0$, $x$
and $y$ satisfy \(log(y) = \alpha - \beta log(x)\).

\subsection{Klienberg-Watts-Strogatz(KWS) small world topology}
Our next network model is inspired by the network of social
contacts. It is well known that any two people are linked by a chain
of half a dozen others who are pairwise acquainted -- known as the
`small-world' phenomenon. This idea was popularised by Milgram in the
60s~\cite{M67}.

The KWS graph topology models a small world network that encapsulates
the following: a network rich in local connections, with a few long
range connections. The network generation starts from a N by N lattice
each point representing an individual in a social network. The lattice
distance \( d((i,j ), (k,l)) =|k-i|+|l-j|\). For a parameter $p$,
every node $u$ has a directed link to every other node $v$ within
$d(u,v) \leq p$. For parameters $q$ and $r$, we construct $q$ long
range directed links from $u$ to a node $v$ with a probability
distribution \([Pr (u,v)]= \frac{(d(u,v))^(-r)}{\sum_{v} (d(u,v))^(-r)
}\).

Low $r$ values means long-range connections, whereas higher values
lead to preferential connections in the vicinity of $u$.

\subsection{LiveJournal (LJ)}
In order to test our ideas on a real world unstructured network, we
turned to a large-scale social network called LiveJournal
(LJ). LiveJournal is a social networking and blogging site with
several million members and a large collection of user defined
communities. LiveJournal allows members to maintain journals,
individual and group blogs, and -- most importantly for our study here
-- it allows people to declare which other members are their
friends. Using a web crawler called touchgraph
(http://www.touchgraph.com), we traced the LJ network to the online
friendship network. The snapshot of the network we use in our analysis
has 3,746,240 nodes and 27,430,000 edges.

A mix server bundled along with a future LiveJournal client acts as
the basis of mix deployment. Mix circuits are built on top of the
social network topology.

\subsection{Expander graphs}
Danezis~\cite{D03} previously analyzed the use of expander graph
topologies to construct mix networks. Expanders are well known to have
excellent expansion properties. We include this as a baseline
comparison against theoretical structured topologies. An expander
graph $G_{N, D}$ has a homogeneous topology with $N$ nodes each with a
degree $D$.

\section{Evaluation framework for measuring traffic analysis resistance}
Before we set out the evaluation framework, we first clarify what we
 mean by ``anonymity'' in this paper. The focus of this work is on
 message anonymity~\cite{SD02}: given a message, the attacker should
 not be able to determine who sent it to whom. There are other
 definitions such as relationship anonymity defined by Pfitzmann
 et. al.~\cite{PH00}.

The objective of our analysis is to determine how the topology of a
 mix network affects the amount of effort on the attacker's part to
 uniquely identify communication endpoints using traffic analysis
 attacks alone. The effectiveness of such attacks depends heavily on
 the topology of the underlying network. If the attacker is not able
 to reduce anonymity beyond his or her initial knowledge then the mix
 network is said to be resistant to traffic analysis attacks under
 the given threat model.

The attacker might also employ side channel analysis on the end-points
before the data enters the mix network, we do not consider such
attacks here. Side channel information might be timestamps or other
information related to the protocol or mechanism in use. Attacks using
such information can be used to link messages to the communication
end-points, and are known as {\em traffic confirmation attacks}
\cite{RSG98}, their effectiveness depends on the mixes' batching and
flushing strategy.

\subsection{Threat Model}
Throughout this paper we consider the adversarial context of a global
passive adversary.

\subsection{Measuring anonymity}
There are several ways one can express the anonymity a system
provides. In our analysis we use a quantitative method due to
Serjantov and Danezis \cite{SD02}, based on the following definition:
``Anonymity of a system may be defined as the amount of information
the attacker is missing to uniquely identify an actor's link to an
action''. In information theoretic terms, the anonymity of the system
$\mathcal{A}$, is the entropy $\mathcal{E}$, of the probability
distribution over all the actors $\alpha_{i}$, that they committed a
specific action.
\begin{equation}
  \mathcal{A}=\mathcal{E}[\alpha_{i}] = -
  \sum_{i}Pr[\alpha_{i}]log_{2}Pr[\alpha_{i}]
  \label{eqn-entropy1}
\end{equation}

This gives the number of bits of information, with a negative sign,
that the attacker is missing before they can uniquely identify a
sender or a receiver.

%In analysing the anonymity of a social network topology, we will
%restrict ourselves to considering {\em traffic analysis attacks}
%alone. The effectiveness of such attacks depends heavily on the
%topology of the underlying network. If the attacker is not able to
%reduce $\mathcal{A}_{network}$ beyond his or her initial knowledge
%then the mix network is said to be resistant to traffic analysis
%attacks under that threat model.

%The attacker might also employ side channel analysis on the end-points
%before the data enters the mix network, we do not consider such
%attacks here. Side channel information might be timestamps or other
%information related to the protocol or mechanism in use. Attacks using
%such information can be used to link messages to the communication
%end-points, and are known as {\em traffic confirmation attacks}
%\cite{RSG98}, their effectiveness depends on the mixes' batching and
%flushing strategy.   
%
\subsection{Modeling mix route selection}
In order to understand the maximal anonymity provided by a mix network
we use Markov chains to model the route selection process, as they
closely match the way mixes are selected to form a mix route.

The process of selecting a mix route of length $k$ by selecting $k$
random nodes in the mix network, is equivalent to first selecting a
random mix node, and, then a random neighbour of the first mix,
repeating this process $k-2$ times. Hence we may model the route
selection process as a random walk on the underlying graph, with the
various states of the Markov chain process being the mix nodes of the
network.

\subsection{Measuring mix network efficiency}
In analyzing the anonymity provided by a particular network topology
we need to examine the probability that a specific message is at a
particular node at a certain time. In order to link the sender and the
receiver to a particular message, the attacker must retrace the steps
taken by the message through the mix network starting from the
receiver. Let the mix network be an undirected graph $G(V, E)$. If
messages $m_{ij}$ are inserted at node $i$ destined for $j$, then for
a message $m_{x}^{t}$ at node $x$ at time $t$, the attacker must link
$m_{x}^{t}$ to $m_{ij}$. Note that $m_{x}^{t}$ might either be in the
edge or the core of the mix network.

Applying the above mentioned information theoretic metric we have:
\[ \mathcal{A}=\mathcal{E}(p_{ij})\] where \(p_{ij}=Pr[m_{x}^{t}\ is\ m_{ij}] \)
 is the probability distribution over all the nodes in $V$.

Suppose a message is inserted into the mix network through a randomly
chosen node. Then after an infinite number of steps, the probability
that the message is present on any randomly chosen node in the network
is given by stationary distribution of the Markov chain $\pi$.  Let
$q^{(0)}$ be the initial probability distribution describing the node
on which message $m$ is introduced into the mix network, this is
equivalent to the distribution of input load across the nodes in
network.  $q^{(t)}$ then, is the probability distribution of the node
on which the message is present after $t$ steps. (this is also known
as the state probability vector of the Markov chain at time $t \geq
0$). With increasing $t$ one would like to see that $q^{(t)}$ merges
with $\pi$. The rate at which this takes place is known as the {\em
  convergence rate} of the Markov chain, and the difference itself is
called the {\em relative point-wise distance} defined as:
\begin{equation}
\Delta(t) = max_{i} \frac{|q_{i}^{t}-\pi_{i}|}{\pi_{i}}
\label{mc-delta-qpi}
\end{equation}

The smaller the relative point-wise distance, faster the convergence,
and more efficient the mix network. It is now easy to see that the
maximum anonymity the network can provide is the entropy of the
stationary distribution of the chain.

\begin{equation}
\mathcal{A}_{network}=\mathcal{E}(\pi)
\label{eqn-max-anon}
\end{equation}.

%A vector $x$ of a matrix $M$ is an eigenvector, with corresponding
%eigenvalue $\lambda$ if and only if \(xM=x\lambda\). 
%
%In simpler terms, this means that the eigenvector is a special vector
%that remains unchanged (or merely linearly scaled) by the application
%of the transformation consisting of multiplication by $M$. The
%eigenvalues of a graph are the eigenvalues of its adjacency matrix,
%and finally the eigenvalues of a Markov chain are the eigenvalues of
%the corresponding transition matrix. 

When $P$ is the transition matrix of the chain it is well known that
$P$ has $n$ real eigen-vectors $\pi_{i}$ and $n$ eigenvalues
$\lambda_{i}$ \cite{W01}.

By using the relation \(q^{(t)}=q^{(0)}P^(t)\), we calculate the
probability distribution of a message being on a node after having
transited a mix route of length $t$.

\subsection{Compromised mixes}
Suppose a subset of mixes are taken over by an adversary. Then a
compromised mix route is defined as a mix circuit that is solely
composed of compromised mix nodes.  Then, what is the probability that
a randomly chosen mix route is compromised?

A network topology with poor expansion properties (or lower
\emph{eigen-value gap} $\epsilon=1-\lambda_{2}$) tends to have
relatively 'localized' mix routes, so that, given the first mix of a
route, there exists a subset of mixes within the network that have a
higher chance of being on the route than others.

The spectral theory of graphs lends us a few tools, namely chernoff
bounds, in quantifying this risk. Suppose $S$ is the set of subverted
nodes, and $\pi_{S}$ the corresponding probability mass of the
stationary distribution $\pi$. The upper bound of the probability that
a mix route (random walk) of length $t$ goes through $t_{S}$ nodes of
$S$ is given by Gilbert~\cite{G98}: \( Pr[t_{A}=t] \leq \left(
1+\frac{(1-\pi(A))\epsilon}{10} \right) e^{-t \frac{(1-\pi(A))^2
\epsilon}{20}}\). However as Danezis~\cite{D03} notes, given that this
probability exponentially decreases with increase in $t$, a small
increase in route length will successfully mitigate this risk.

What is more relevant in the context of unstructured networks, is the
presence of 'hub' nodes and 'weak-ties'. Hubs~\cite{N03} are
special nodes that owing to their position in the network topology
handle large amounts of traffic reducing. Similarly,
weak-ties~\cite{G73} are edges responsible for significantly
reducing average path-lengths in networks of tightly knit communities
such as social networks.  The risk of compromised mix routes is
significantly higher in a topology where hubs only connect to other
hubs, and handle most of the network traffic. If an attacker can
locate and strategically target mix nodes that also play the role of a
hub, then the percentage of mix routes under risk can be significant.

Hence, we simulated a large number of random walks for various
topologies presented in section~\ref{sec-netmod}, of different
lengths, and make a recommendation on the route length to mitigate
this risk in section~\ref{sec-res-rl}.

\subsection{Intersection attacks}
The term {\em intersection attack} was introduced by Berthold
et.al.~\cite{BPS00}. These attacks involve the detection of the
preferential use of a mix route. If for some reason, a sender under
attack sends more traffic along a specific route much more often than
other routes, then a simple intersection attack is carried out by
intersecting the set of possible next-hop mixes of every mix with the
set of possible next-hop destinations of previous messages. The the
actual path of a message will then become apparent unless the network
has countermeasures against observability.

If each link from a mix node is used to flush messages to its
neighbours, then the potential for the simplest of intersection
attacks can be greatly reduced~\cite{KAP02}. So, for a given node $i$,
we wish to calculate the probability that any out going link remains
unused during a flushing cycle. If each mix node receives $b$ messages
per batch, then each of these will appear on a particular outgoing
link $j$ with a binomial probability distribution
$p_{i}=1/deg_{i}$. Danezis~\cite{D03} then calculates the volume of
incoming traffic required so that the probability of any out going
link being unused is negligible.

\begin{equation}
b=\frac{9}{f^{2}} \left( \frac{1-p_{i}}{p_{i}} \right)
\label{intsc-attack}
\end{equation}
where $f$ is the percentage deviation of traffic output on a
particular link of $i$ in a given flushing cycle from the mean traffic
output.

Combining this with $p_{min}$, the probability associated with the
highest degree node in the mix network, we can derive the amount of
genuine traffic to be mixed together. 

The prevention of basic intersection attacks as a system design
criteria can be traced back to the work of Reiter and
Rubin~\cite{RR98}.

\section{Results and Discussion}

\subsection{Simulation parameters}
\label{sec-res-rl}
In all the synthetically generated networks we considered, we $N
\approxeq 5000$ nodes. The parameters used for each of them are listed
below.

We model scale-free networks with linear preferential attachment with
$m$ links per node and average node degree $\langle d \rangle$ ; $2
\leq m \leq 7$ and $4 \leq \langle d \rangle \leq 14$.

Next we model scale-free random networks which have a scale-free
degree sequence but which are random in all other respects. Generated
with parameters $\alpha=0.25$, $\beta=0.25$ and Average node degree $4
\leq \langle d \rangle \leq 14$. See section~\ref{sec-sfrgt} for an
explanation of $\alpha$ and $\beta$

Klienberg-Watts-Strogatz model of directed social network ties is
analyzed next, generated with parameters $r$, the lattice radius
within which each node creates direct links to all its neighbors. $q$
is the number of weak ties. We used $1 \leq r \leq 4$ and $2 \leq q
\leq 10$.

Our next network is based on our primary source data, obtained by
web-crawling the LiveJournal site. The snapshot of the network we use
in our analysis has 3,746,240 nodes and 27,430,000 edges.

Finally we analyze two theoretical topologies, one degree heterogeneous
and the other degree homogeneous, to offer a baseline comparison
against ER graph and constant expander graph topologies.

The ER graph is created with each edge formation as the result of a
Poisson distribution of $p=0.0028$ with $\langle d \rangle = 14$.

The constant expander graph is created with each node having $D=14$
edges. Motwani et.al.~\cite{MR95} prove a relation between the second
eigenvalue $\lambda_{2}$ of the transition matrix of a constant
expander graph and the degree $D$ of a node \( \lambda_{2} \geq
\frac{2\sqrt{D-1}}{D}\). We can then use the result of
Sinclair~\cite{S93} connecting $\lambda_{2}$, random walk length $t$
and convergence rate $\Delta(t)$, namely \(\Delta(t) \leq
\frac{\lambda_{2}^{t}}{\min_{i \in V} \pi_{i} }\). For $D=14$, we have
a constant expander graph with theoretical minimum second eigen-value
of $\lambda_{2} \geq 0.5527708$, converging to maximal anonymity state
in approximately 4 steps. This forms the baseline against which we
compare all the other topologies.

\subsection{Efficiency}
We can now comment on the efficiency and recommended mix route lengths
for various network topologies by comparing them to our baselines.

\begin{figure}
\begin{minipage}[t]{6cm}
\vspace{0pt}\includegraphics[width=\textwidth]{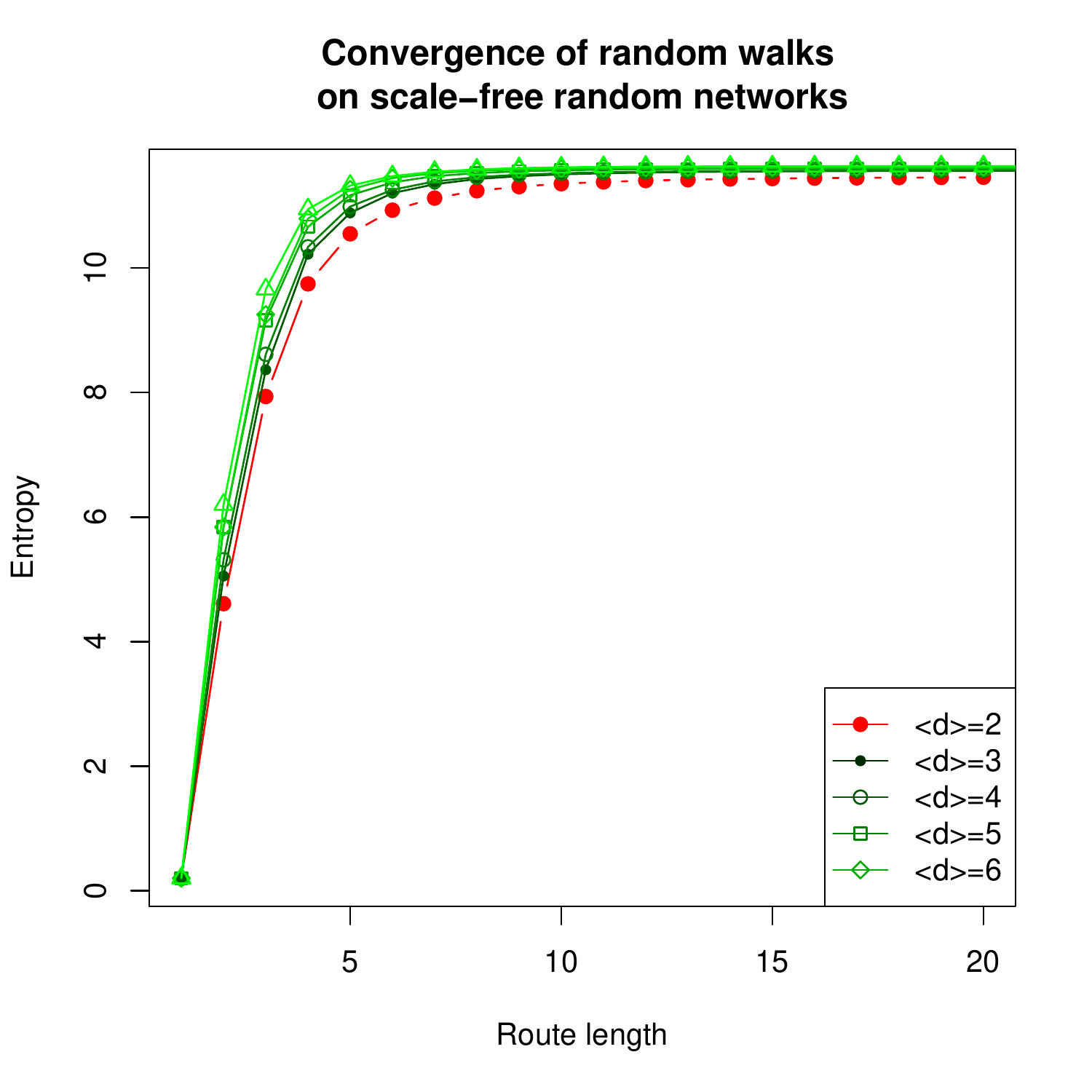}
\end{minipage}
\begin{minipage}[t]{5cm}
\vspace{2pt}{\begin{tabular}{@{}lccc@{}}\toprule
      Network ($N=5000$) & $\langle d \rangle or D$ &  $t$ & $\mathcal{A}_{network}$\\ \midrule
      SFR     & 2                     &  ~8  & 11.4383\\
              & 3                     &  ~7  & 11.5626\\
              & 4                     &  ~6  & 11.5958\\
              & 5                     &  ~6  & 11.6135\\
              & 6                     &  ~5  & 11.6351\\
      \addlinespace
      ER      & 14                    &   7  & 12.2339\\
      \addlinespace
      Expander& 14                    & 4    & 12.2877\\ \bottomrule
    \end{tabular}}
\end{minipage}
\caption{Convergence rates: Efficiency and maximal anonymity for
Scale-free random, ER and Constant expander graph topologies}
\label{entropy_sfr}
\end{figure}

The efficiency of mix topologies based on a scale-free random networks
 is shown in Figure~\ref{entropy_sfr}. It plots the anonymity achieved
 against increasing random walk lengths. Maximal anonymity is
 calculated using equation~\ref{eqn-max-anon} is the entropy of the
 probability distribution of the chain at convergence.

Our calculations show that maximal anonymity is reached in just 6
 steps in the medium density case $\langle d \rangle \geq 4$, as
 opposed to 4 steps in to 4 steps in an expander graph topology.  It turns out that
 social collaboration networks~\cite{N01a,N01b,N01c} with scale-free
 characteristics have average degrees in the range of $4 \leq \langle
 d \rangle \leq 18$. This suggests, firstly, that efficient mix
 networks can be designed using scale-free random networks, and
 second, that mildly denser scale-free networks are more suitable for
 building mix networks than sparser ones.
 
While this is an encouraging initial result, it is important to strike
a note of caution. Scale-free random graphs only model the scale-free
aspect of degree distribution, while being random in every other
way. However most real world unstructured networks have several other
non-random characteristics apart from their degree
distributions. 

\begin{figure}
\subfloat[Klienberg-Watts-Strogatz model]
{ \includegraphics[width=0.45\textwidth]{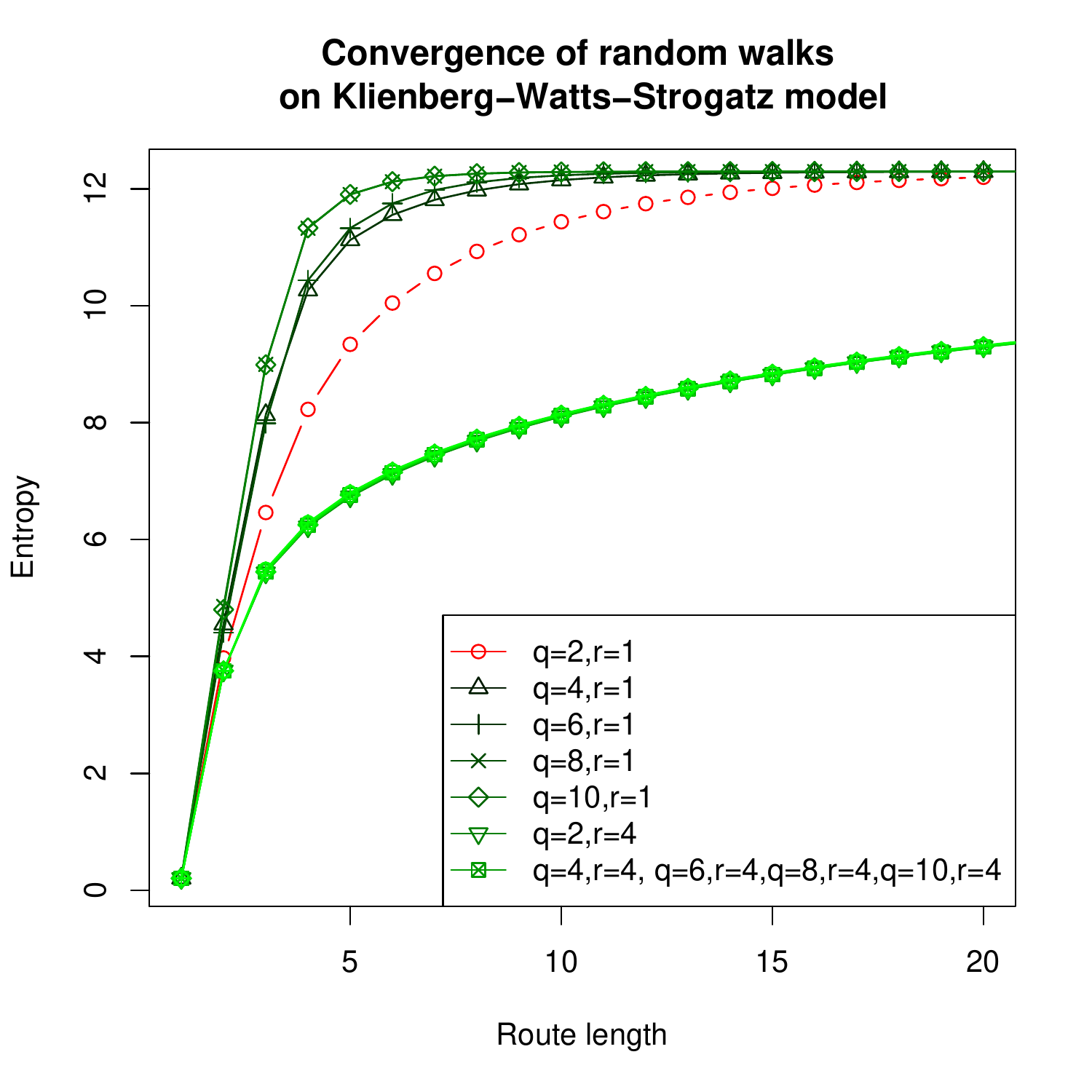}}
\hfill
\subfloat[Scale-free network with preferential attachment]
{\includegraphics[width=0.45\textwidth]{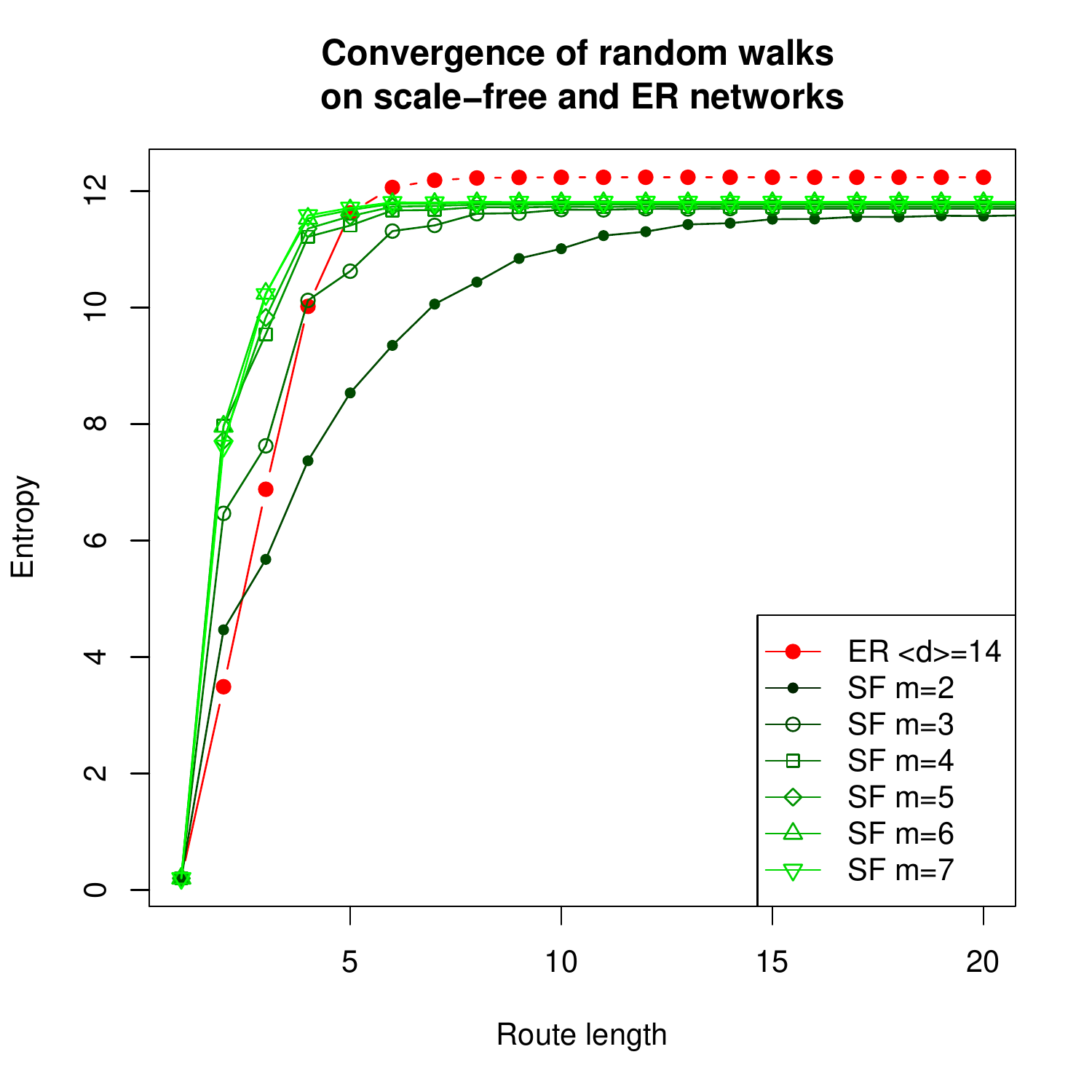}}
\caption{Mean entropy vs mix-route length}
\label{entropy_ws_sf}
\end{figure}

A number of real world unstructured networks are not scale-free, hence
we included the Klienberg-Watts-Strogatz(KWS) network topology, as it
explicitly models the presence of weak ties in a network. We
experiment with a number of parameter configurations; selecting $r=1$
and $r=4$ to model low and high richness in local links or 'strong
ties' between nodes; and $2 \leq q \leq 10$ the number of short cuts
or 'weak ties', between mix nodes. Figure~\ref{entropy_ws_sf}-a plots
mix-route length vs mix network anonymity, for the KWS topology. When
the topology is poor in local links, it seems to converge in 7 to 8
hops, given enough short cuts. However, if the network invests a large
amount of resources into local connections forming relatively tightly
knit communities, then regardless of the amount of shortcuts,
convergence is not achieved until 62 hops!

Our final model network topology is the scale-free network based on
linear preferential attachment, which has attracted much attention in
the complex networks literature. This topology models a scale-free
network where hubs are connected to other hubs, a pattern that is
repeatedly observed in many real world scale-free networks. The
parameter $m$ controls graph sparsity, random walk and convergence
results are shown in figure~\ref{entropy_ws_sf}-b. Our simulations
show that while very sparse topologies converge in 10 to 15 hops,
topologies that are relatively dense converge within 6 or so hops,
this is comparable to the optimal 4 hops of a constant expander graph.

\begin{figure}
\begin{minipage}[t]{6cm}
\vspace{0pt}\includegraphics[width=\textwidth]{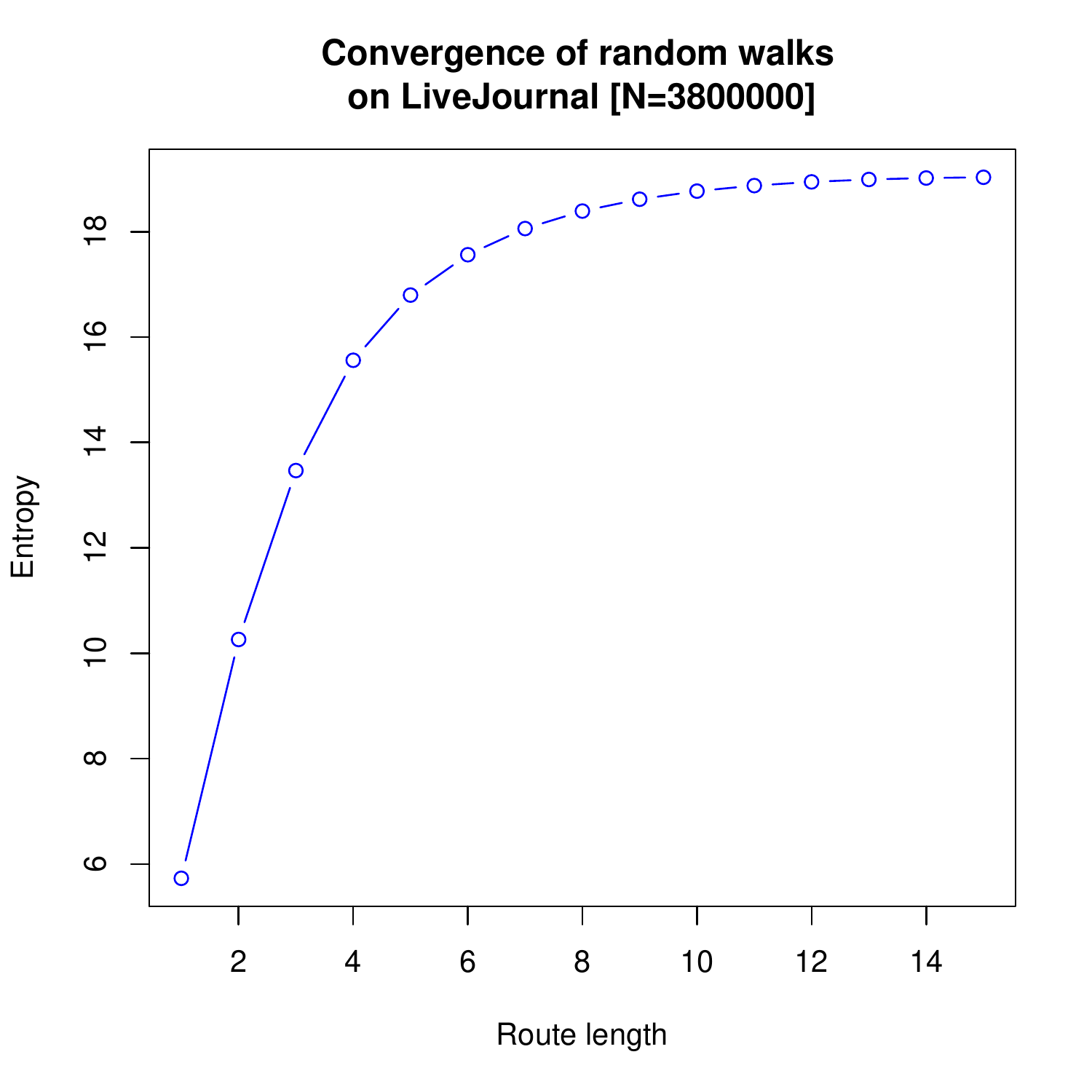}
\caption{Mean entropy vs mix-route length for LiveJournal}
\label{entropy_lj}
\end{minipage}
\hspace{2em}
\begin{minipage}[t]{5cm}
\vspace{0pt}{\begin{tabular}{@{}lccc@{}}\toprule
      SF & $m$ &  $t$ & $\mathcal{A}_{network}$\\ \midrule
         &  2  &  15 &  11.5852 \\
         &  3  &  10 &  11.6961 \\
         &  4  &  6  &  11.7293 \\
         &  5  &  6  &  11.7687 \\ 
         &  6  &  6  &  11.7953 \\ 
         &  7  &  6  &  11.8090 \\ 
     KWS & $q=2,p=1$ &  11      &  12.2945 \\
         & $q=10,p=1$&  5      &  12.2939 \\
         & $q=2,p=4$ &  63      &  11.6440 \\
         & $q=10,p=4$&  63      &  11.6380 \\
      \addlinespace
      ER      & $\langle d \rangle =14$  &   7  & 12.2339\\
      \addlinespace
      Expander& $D=14$                  & 4    & 12.2877\\ \bottomrule
    \end{tabular}}
\caption{Convergence rates: Efficiency and maximal anonymity for
linear Scale-free, KWS, ER and constant expander graph topologies}
\label{summary-sf-kws}
\end{minipage}

\end{figure}

Next, we considered our primary data source the LiveJournal graph with
a little less than 4 million nodes. Figure~\ref{entropy_lj} shows the
convergence rate of mix routes, which we note converges to the
stationary distribution in around 11 hops. While this seems a high
number in comparison to expander graphs (converging in 4 hops), we
also note that the entropy achieved by the random walk in 4 hops in LJ
is $\mathcal{A}_{LJ}^{4}= 15.56$. To obtain the equivalent on an
expander graph topology we would only need $2^{15.56}$ or 48309
nodes. On the face of it the design decision seems really simple, to
go for a structured expander graph topology. We argue a different
view: A successful mix network design must also consider liability
management issues arising from running a mix. Considering that aspect,
topology links backed up by social capital are likely to be more
robust than those of an optimal topology, but where nodes quickly
buckle under legal pressure. We propose, running mixes on the nodes of
the LJ topology bundled along with a future LiveJournal client. Nodes
only allow incoming traffic from their neighbors, and will only direct
outgoing traffic to their neighbors. 

In this context, an interesting question is why nodes would process
traffic that didn't originate from their neighbors, and especially so
in the face of legal hassles?

We offer the following reasoning: Humans making decisions on whether
or not to run a mix server, will have to consider the following
costs. They benefit in the long term, from processing traffic for
unknown nodes in order to generate a diverse user base, the need for
which is well illustrated in Dingledine and
Mathewson~\cite{DM06}. However this only holds if other mixes
cooperative accordingly. Then there is the immediate social benefit of
having processed traffic for your friends. The success of the system
then depends on the extent to which individual nodes perceive the
costs of litigation pressure to be less than the total of immediate
social benefit and the long term benefit of a diverse user
base. Psychology studies tell us that humans involved in taking
security decisions weigh short and long term benefits differently. It
should also be interesting to investigate whether the idea of running
a mix to primarily process traffic for your friends is an effective
tool for seeding indirect reciprocity in a mix network where
cooperation flourishes.

\subsection{Compromised mix nodes}
As explained in our evaluation framework, compromised nodes can lead
to compromised routes. This presents a special challenge in
unstructured networks where $\pi_{A}$, the probability mass of the
stationary distribution $\pi$, corresponding to set of compromised
nodes $A$, can be significant for topological reasons.

To measure the robustness to nodes being strategically compromised by
an attacker, we simulated 100000 random walks of different lengths,
for each of our network topologies, in the range indicated by
efficiency considerations of the previous section $3 \leq t \leq 6$,
and measured the fraction that passed through compromised nodes.  The
set of compromised nodes is chosen to consist of the nodes with the
highest degrees in the network.  In each case, for mix routes greater
than 4 hops the probability of existence of a compromised mix route is
negligible. Fig~\ref{corruptnodes} in the appendix confirms that the
threat of mix route compromise can be successfully reduced by suitably
increasing the mix-route length.

\subsection{Intersection attacks}
Using equation~\ref{intsc-attack} we consider the required batch sizes
for a threshold mix, so that the traffic output on any link in the mix
network does not deviate by more than 5\% from the mean traffic output
on that link. For $f=5$ we calculate the number of messages that must
be received in each mixing cycle in table~\ref{summary-intsc-att}.

\begin{table}
\centering
\begin{tabular}{@{\extracolsep{1em}}llccc@{}} \toprule
Network    & $\langle d \rangle$ & $p_{min}$    &  Batch size\\\midrule
SFR        & 2  &     0.0344     & 10.08\\
           & 3  &     0.0222     & 15.84\\
           & 4  &     0.0243     & 14.4\\
           & 5  &     0.0192     & 18.36\\ 
           & 6  &     0.0135     & 26.28\\ 
           & 7  &     0.0125     & 28.44\\
\addlinespace
KWS        &   27 ($q=1,r=1$)   &     0.0294     &  11.88\\
           &   43 ($q=10,r=1$)  &     0.0169     &  20.88\\ 
           &   26 ($q=1,r=4$)   &     0.0333     &  10.44\\ 
           &   28 ($q=10,r=4$)  &     0.0294     &  11.88\\ 
\addlinespace
SF-linear  & 4 &     0.0048     &  74.16\\ 
           & 6 &     0.0048     &  74.16\\
           & 8 &     0.0041     &  86.04\\ 
           & 10 &    0.0038     &  93.6\\
           & 12 &    0.0037     &  96.12\\ 
           & 14 &    0.0031     &  112.32\\
\addlinespace
LJ           & 7.3221 & 0.00857  & 41.64\\
ER           & 14      & 0.0333 & 10.44\\
Expander     & 14      & 0.0714 & 4.68\\  \bottomrule
\end{tabular}
\caption{Batch sizes required to prevent intersection attacks}
\label{summary-intsc-att}
\end{table}

From table~\ref{summary-intsc-att} it is clear that scale-free random
networks and KWS both require a batch size that is 4-5 times that of
expander graphs. Whether social networks can produce enough 'chatter'
to feed genuine traffic into the mix network is an open question.

Our theoretical base line of ER network topology does slightly better
at a little over twice that. More significantly, the LJ network has a
batch size of almost 9 times the required batch size for expander
graphs. Scale-free networks with linear preferential attachment are
the worst performing, requiring a batch size almost 20 times larger
than expanders. We think that the exceptionally high value of batch
size in LJ network is due to its size of four million or so
nodes. While does not mean that LJ is inherently unsuitable as a mix
network topology, but it certainly indicates a scalability limit with
the deployment of mixes on LJ nodes, as proposed earlier.

%From the above, we conclude that the batch sizes in the case of
%scale-free networks need to be significantly larger than that of
%expander topologies for the same intersection attack threat. Hence,
%the link capacity will have to be substantially larger than in
%expander graphs and could pose a difficult challenge to meet in some
%cases.

%\section{Future Work}
%Future directions involve obtaining analytical bounds on the mixing
%rates of the Klienberg-Watts-Strogatz and the random scale-free model
%using path coupling methods.
%
%Additionally, the risk of mix-route corruption needs to be examined
%for other topology attack such as the Freeman betweenness centrality
%and Bonacich eigen-vector centrality attacks.
%
%The authors are in the process of expanding the analysis to compare
%de Bruijn graph topologies with unstructured networks.
%
%Another important direction is to carry out a comprehensive analysis
%of running mix networks over unstructured topologies in a low-latency
%setting.
%
%The probability that a path is totally controlled by compromised nodes
%depends on the amount of traffic processed by the compromised nodes,
%and the expansion properties of the graph. It should therefore be
%interesting to analyze the threat posed by adversaries that compromise
%nodes based on graph centrality algorithms developed as part of the
%social network analysis literature \cite{WF94}, and possible counter
%measures.
%

\section{Conclusions}
We have analyzed a comprehensive set of network topologies from the
perspective of efficiency, maximal anonymity, compromised nodes and
simple intersection attacks in comparison with (provably optimal)
expander graphs. 

To the standard threat model of the global passive adversary, we have
added real world issues such as liability management and the
need for clear incentives for carrying traffic under the pressure of
legal threats, and discussed our simulation results in this context.

We considered topologies with two important characteristics found in
empirical studies of large-scale unstructured networks: scale-freeness
(scale-free random graph) and the small-world property
(Klienberg-Watts-Strogatz (KWS) graph). In both the topologies, we can
recommend mix route lengths for achieving 95\% of maximal anonymity,
that is only a few hops larger than the optimal route length found in
expander graph topologies. Currently deployed mix networks such as TOR
have around 540 volunteers. To increase the scale of such mix
deployments the Internet, we believe the way forward (for high latency
systems only) is to use online social networks. The minimum mix route
must have three mixes to allow sender and receiver anonymity. For this
length, a mix network constructed by placing mixes on the nodes of a
social network such as LiveJournal can achieve far higher maximal
anonymity as per the entropy metric we have used. We argue that
including network incentives within a framework does not allow the
construction of structured overlay mix topologies that can robustly
withstand the threat of legal action. By moving to social networks, we
make a start on tapping the social capital underlying node-node
interaction to encourage users to deploy and run mixes with policies
that reflect this aspect.

We also found that subverted nodes, either compromised randomly, or by
strategic choice, on the basis of their degrees has little effect on
the efficiency of a mix network. This is because the route length
required to mitigate that risk is less than the recommended length for
achieving efficient convergence rates.

We also analyzed scale-free and the small-world topologies for their
robustness to attacks based on traffic load patterns observable on
their out-going and in-coming links. Both the scale-free random graph
topology and the KWS topology turn out to require almost 5 times as
much traffic as corresponding expander graph topology. This suggests
the need for further tests to see if enough genuine traffic is
generated in online social network interaction, to satisfy the minimum
batch sizes required for preventing the most basic versions of these
attacks.

We conclude that, unstructured networks based on large-scale
topologies are indeed very promising, we have outlined the merits and
challenges these topologies present to the design of mix networks for
anonymous communication.

\section{Acknowledgements}
The authors are grateful to Ross Anderson for reviews on early
versions of the paper, and to George Danezis and Roger
Dingledine, for thought provoking discussions.

\appendix
\section{Mix-route compromise on linear preferential attachment scale-free networks}
\label{app-corrupt-route}
In this section we sketch a few analytical results concerning
mix-route compromise in BA scale-free networks.

Let $B$ be the set of compromised high vertex-order centrality
nodes. For a route to be fully compromised, all intermediate nodes
must be in $B$. We then wish to calculate, 

\(P(C_{l})=[Pr (Random-Walk(v_{1}....v_{l}))] \forall {v_{1}...v_{l}}
\subseteq B\).

It is straightforward to see that if \(l>|B|\) then P(C)=0.  In BA
scale-free networks, all hubs(high vertex-order) nodes are connected
to each other. Hence, 

\[P(C)=\frac{|B|-1}{\prod_{{\scriptscriptstyle j \in B}} k_{j}}\]

%\bibitem{NL} National Laboratory for Applied Network Research, Routing
%  Data, http://moat.lanr.net/Routing/rawdata.

\begin{figure}[htbp]
%\begin{minipage}{7cm}
%\vspace{0pt}\includegraphics[width=0.45\textwidth]{data/corrupt_nodes_sim-1.pdf}
%\end{minipage}
%\begin{minipage}{7cm}
%\vspace{0pt}\includegraphics[width=0.45\textwidth]{data/corrupt_nodes_sim-2.pdf}
%\end{minipage}
%\begin{minipage}{7cm}
%\vspace{0pt}\includegraphics[width=0.45\textwidth]{data/corrupt_nodes_sim-3.pdf}
%\end{minipage}
%\begin{minipage}{7cm}
%\vspace{0pt}\includegraphics[width=0.45\textwidth]{data/corrupt_nodes_sim-4.pdf}
%\end{minipage}
%%
%
\subfloat[length=$3$]{  \includegraphics[width=0.45\textwidth]{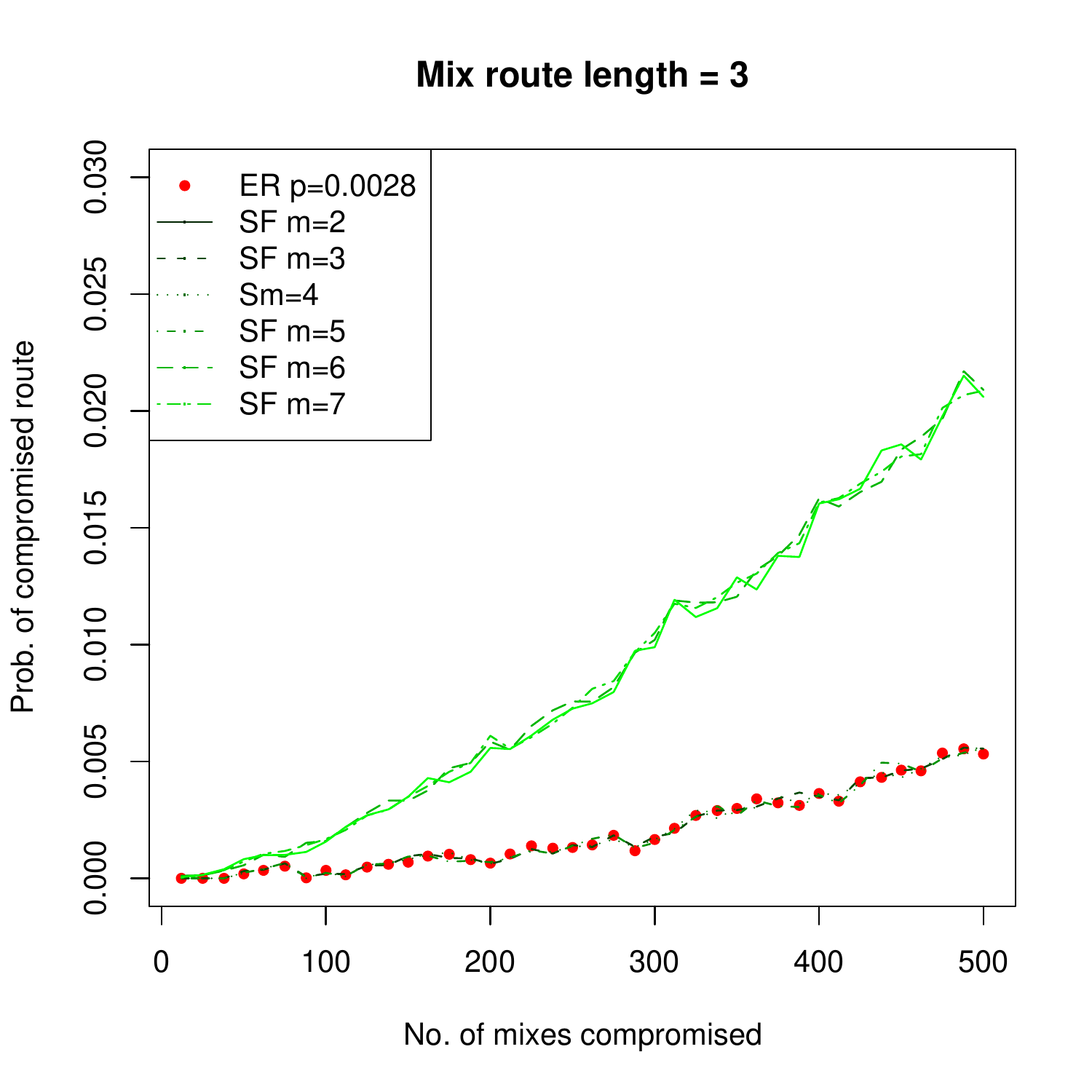}}
\subfloat[length=$4$]{  \includegraphics[width=0.45\textwidth]{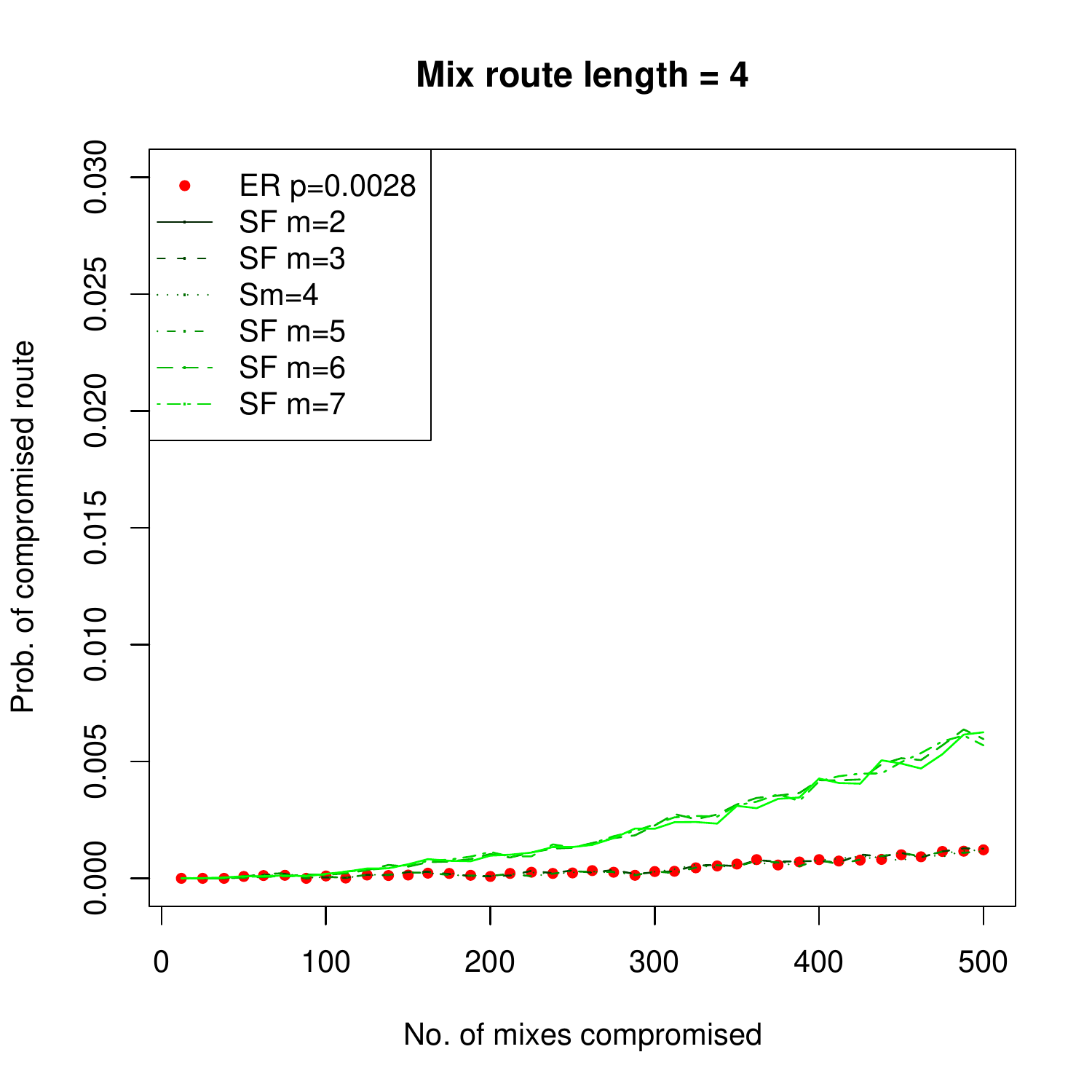}}
\subfloat[length=$5$]{  \includegraphics[width=0.45\textwidth]{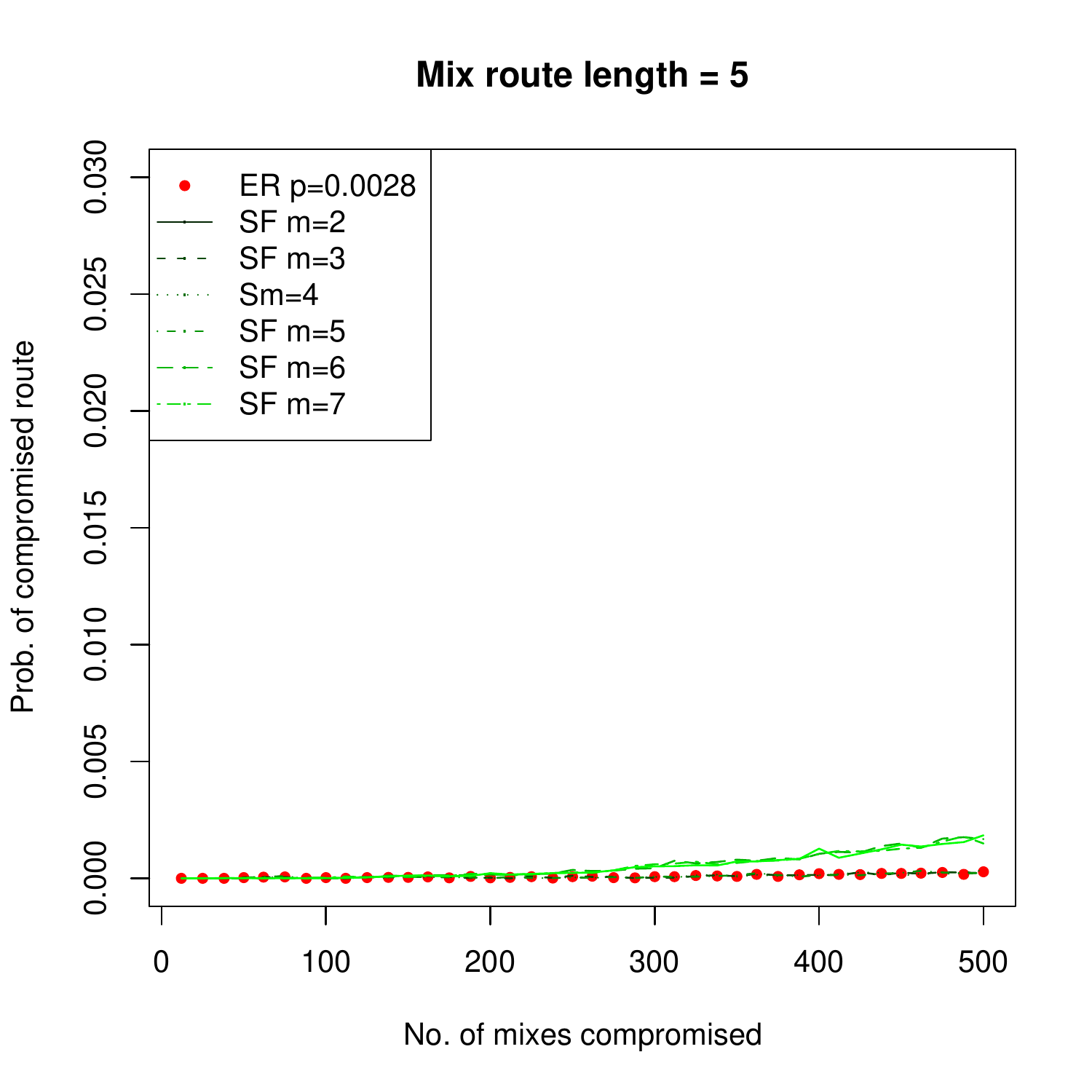}}
\hfill
\subfloat[length=$6$]{  \includegraphics[width=0.45\textwidth]{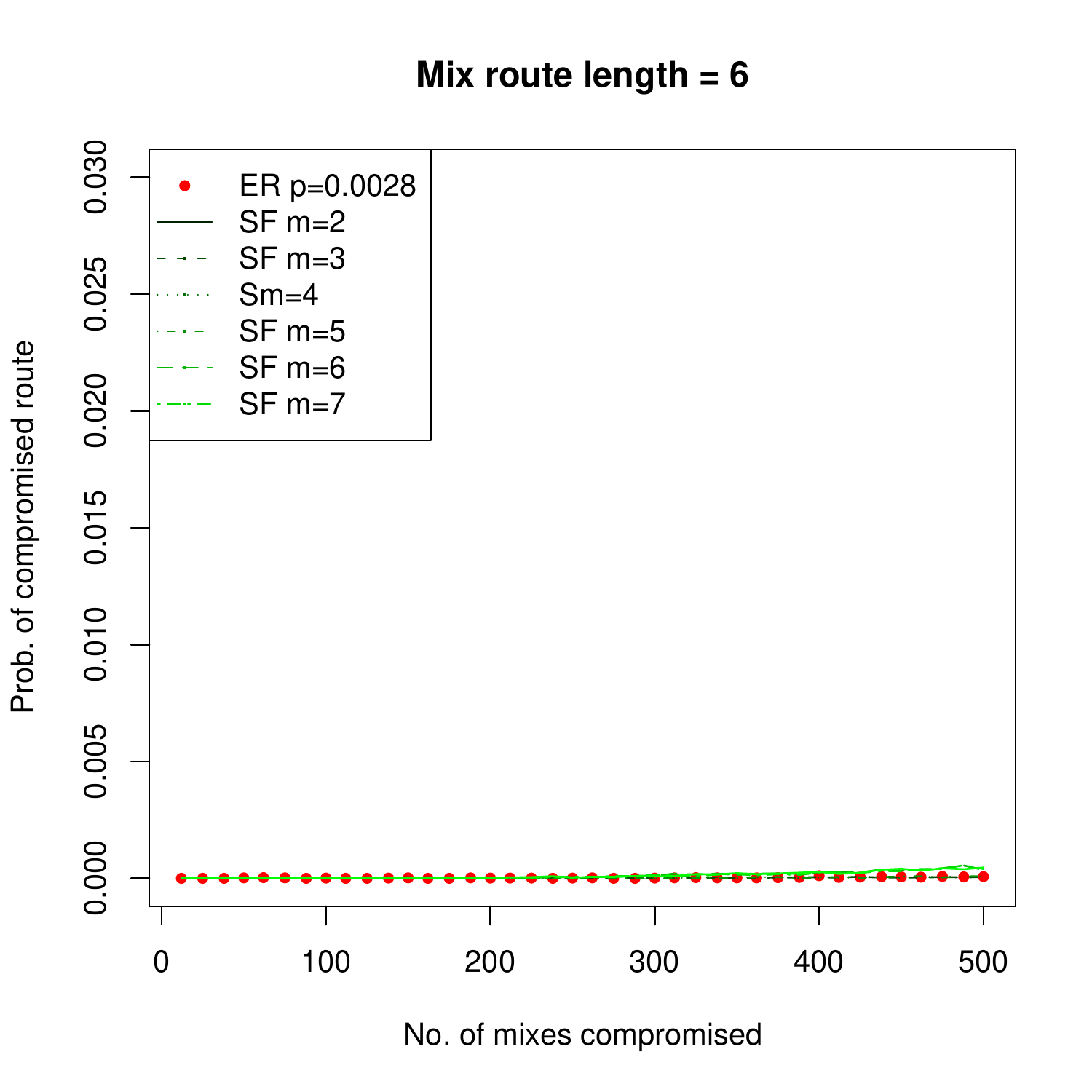}}
\caption{Probability of mix-route compromise vs no. of corrupt nodes}
\label{corruptnodes}
\end{figure}

\section{Convergence rate and network size in scale-free random networks }
Simulations conducted in this paper have not accounted for the effect
of varying network size on the convergence rate of the respective
topologies. We address this, by offering a simple conductance based
proof that the second eigen-value of a scale-free network is a
independent of the network size. See \cite{R06} for a review of the
conductance based technique as well as others.

We denote a scale-free network generated by preferential attachment,
by $G_{m, n}(V, E)$ where $m$ is the number of initial nodes created
at time=$t_{0}$ and $n$ is the total number of nodes in the
network. At every time step $t_{i}, i \ge 0$, $m$ nodes are added to
the networks. For every node added, we create $m$ edges from the node
to existing nodes in the network. We continue until $|V| = n$.

Next, there is an intimate relationship between the rate of
convergence and a certain structural property called the {\em
conductance} of the underlying graph. Consider a randomly chosen
sub-graph $S$ of $G(V, E)$. Suppose a random walk on the graph visits
node $i$ $i \in S$. What is the probability that the walk exits $S$ in
a single hop. If conductance is small, then a walk would tend to ``get
stuck'' in $S$, whereas if conductance is large it easily ``flows''
out of $S$.

Formally, for $S \subset G$, the {\em volume} of S is
\(vol_{G}(S)=\sum_{u \in S}d_{G}(u)\), where $d_{G}(u)$ is the degree
of node $u$. The {\em cutset} of $S$, \(C_{G}(S, \overline{S})\), is
the multiset of edges with one endpoint in S and the other endpoint in
$\overline{S}$. The textbook definition of conductance $\Phi_{G}$ of
the graph $G$ is the following:

\begin{equation}
\Phi_{G}=\min_{S \subset V, vol_{G}(S) \leq vol_{G}(V)/2}
\frac{|C_{G}(S, \overline{S})|}{vol_{G}(S)}
\label{eqn-phi-def}
\end{equation}

\cite{MPS04} prove that the conductance of a scale-free network is a
     {\em constant}. Specifically, \(\forall m \geq 2\ and\
     c<2(d-1)-1\), \( \exists \alpha=\alpha(d,c)\) such that
\begin{equation}     
\Phi=\frac{\alpha}{m+\alpha} \label{eq:1}
\label{eqn-phi-const}
\end{equation}

From \cite{S93} we have the following bound for $\lambda_{2}$:

\begin{equation}
1-2\Phi \leq \lambda_{2} \leq 1 - \Phi^{2}/ 2
\label{eqn-phi-l2}
\end{equation}

Substituting for $\Phi$ from equation~\ref{eqn-phi-const} in
equation~\ref{eqn-phi-l2}, it is easy to see that $\lambda_{2}$ is a
constant.

\bibliographystyle{alpha} \bibliography{unstructured-mixes}

\end{document}